\documentclass[sigconf]{acmart}
\AtBeginDocument{%
  }

\copyrightyear{2023}
\acmYear{2023}
\setcopyright{acmlicensed}\acmConference[WWW '23]{Proceedings of the ACM Web Conference 2023}{April 30-May 4, 2023}{Austin, TX, USA}
\acmBooktitle{Proceedings of the ACM Web Conference 2023 (WWW '23), April 30-May 4, 2023, Austin, TX, USA}
\acmDOI{10.1145/3543507.3583865}

\usepackage{xcolor}
\usepackage{booktabs} 

\renewcommand\arraystretch{1.2}
\usepackage{subcaption}
\usepackage{graphicx}  
\newcommand{\KK}[1]{{}}

\newcommand{\MB}[1]{{}}

\newcommand{\JD}[1]{{}}

\begin{document}

\title{Moral Narratives Around the Vaccination Debate on Facebook}

\author{Mariano G. Beir\'{o}}
\email{mbeiro@fi.uba.ar}
\orcid{0000-0002-5474-0309}
\affiliation{%
  \institution{Universidad de Buenos Aires. Facultad de Ingenier\'{i}a}
  \city{Buenos Aires}
  \country{Argentina}
 }
 \affiliation{%
  \institution{CONICET-Universidad de Buenos Aires.
  INTECIN}
    \city{Buenos Aires}
  \country{Argentina}
 }

 \author{Jacopo D' Ignazi}
 \email{jacopo.dignazi@isi.it}
\affiliation{%
 \institution{ISI Foundation}
 \city{Turin}
 \country{Italy}}

\author{Mar\'{i}a Florencia Prado}
\affiliation{%
  \institution{Universidad de Buenos Aires. Facultad de Ingenier\'{i}a}
  \city{Buenos Aires}
 \country{Argentina}
 }

\author{Victoria P\'{e}rez Bustos}
\affiliation{%
  \institution{Universidad de Buenos Aires. Facultad de Ingenier\'{i}a}
  \city{Buenos Aires}
 \country{Argentina}
 }

\author{Kyriaki Kalimeri}
\orcid{0000-0001-8068-5916}
\email{kyriaki.kalimeri@isi.it}
\affiliation{%
 \institution{ISI Foundation}
 \city{Turin}
 \country{Italy}}

\renewcommand{\shortauthors}{Beir\'{o} M.G. et al.}

\begin{abstract}

Vaccine hesitancy is a complex issue with psychological, cultural, and even societal factors entangled in the decision-making process. 
The narrative around this process is captured in our everyday interactions; social media data offer a direct and spontaneous view of peoples' argumentation. 
Here, we analysed more than 500,000 public posts and comments from Facebook Pages dedicated to the topic of vaccination to study the role of moral values and, in particular, the understudied role of the Liberty moral foundation from the actual user-generated text. 
We operationalise morality by employing the Moral Foundations Theory, while our proposed framework is based on recurrent neural network classifiers with a short memory and entity linking information.
Our findings show that the principal moral narratives around the vaccination debate focus on the values of Liberty, Care, and Authority. 
Vaccine advocates urge compliance with the authorities as prosocial behaviour to protect society. 
On the other hand, vaccine sceptics mainly build their narrative around the value of Liberty, advocating for the right to choose freely whether to adhere or not to the vaccination. 
We contribute to the automatic understanding of vaccine hesitancy drivers emerging from user-generated text, providing concrete insights into the moral framing around vaccination decision-making. 
Especially in emergencies such as the Covid-19 pandemic, contrary to traditional surveys, these insights can be provided contemporary to the event, helping policymakers craft communication campaigns that adequately address the concerns of the hesitant population.

\end{abstract}




\maketitle

Vaccines have saved countless lives; among the significant achievements of vaccines is the eradication of smallpox in the 1960s and 1970s, a disease estimated to have killed up to 300 million people, including six European monarchs, in the 20th-century alone~\cite{henderson2011eradication}.
Despite being the most successful and cost-effective health intervention in human history, vaccine hesitancy continues to grow in both low- and high-income countries \cite{larson2011addressing,salmon2015vaccine}. 
In 2019 the World Health Organization (WHO) declared vaccine hesitancy as one of the top 10 threats to global health~\cite{who2019}.

More recently, a debate mainly concerned parents arguing about children's vaccination~\cite{betti2021detecting} scaled up with the COVID-19 pandemic to a broader point of argumentation~\cite{mejova2020covid}.
Despite the scientific consensus about the safety of vaccines, objections continue to rise with familiar narratives that mainly evolve around science denial~\cite{browne2015going}, conspiracy theories~\cite{jolley2014effects}, and alternative health practices~\cite{kalimeri2019human,kata2010postmodern}.
The drivers of vaccine hesitancy are complex, ranging from the individual (i.e. psychological) to interpersonal (social cycle interactions) and public policy (i.e. national and local agencies)~\cite{DUBE2016,macdonald2015vaccine}.
Not all factors received the same attention from researchers and practitioners; still, there is very little work examining the psychological and, in particular, the moral values underpinning objection towards vaccination.
Even if noted moral reasoning was related to vaccination opinion formation~\cite{kunda1990case},
it was the development of the \textit{Moral Foundations Theory} (MFT) ~\cite{Haidt2004} to provide scientists with a concrete framework to assess moral values systematically.

MFT focuses on the explanation of the psychological basis of morality, its origins, development, and cultural variations, 
and identifying the following five moral foundations \cite{Haidt2007,Haidt2004}; \textit{Care/Harm}: basic concerns for the suffering of others, including virtues of caring and compassion, \textit{Fairness/Cheating}: concerns about unfair treatment, inequality, and more abstract notions of justice,
\textit{Loyalty/Betrayal}: concerns related to obligations of group membership, such as loyalty, self-sacrifice and vigilance against betrayal, \textit{Authority/Subversion}: concerns related to social order and the obligations of hierarchical relationships
such as obedience, respect, and proper role fulfilment, and  \textit{Purity/Degradation}: concerns about physical and spiritual contagion, including virtues of chastity, wholesomeness and control of desires.
More recently, the theory was enhanced with a sixth dimension: \textit{Liberty/Oppression}~\cite{haidt2012righteous}. Liberty expresses the feelings of reactance and resentment towards oppressors. 

Importantly, there is a scientific consensus on the moral viewpoints related to vaccine hesitancy~\cite{amin2017association,kalimeri2019human,rossen2019accepters}.
The common denominator of the argumentation narrative against vaccine uptake includes arguments of ``Purity'' of the body and soul, advocating for alternative health practices, notions of ``Liberty'', intended as freedom of choice as an alternative to obligatory vaccination schemes~\cite{betti2021detecting,kalimeri2019human}.
Till now, most studies have been carried out employing questionnaires which, even if administered online, still have a limited reach.
Here, we assess the moral narratives of vaccine advocates and sceptics on social media (SM), particularly on posts and comments on Facebook Pages related to vaccination.
Such an approach can provide additional sources of information, especially from under-represented countries and populations in the peer-reviewed literature~\cite{larson2014understanding}.

Here, we assess the moral viewpoints of people analysing user-generated text on social media. Focusing on the linguistic measurements, we trained a Long Short-Term Memory (LSTM) deep neural network to infer the expression of moral values in text~\cite{hochreiter1997long}.
This study aims to provide insights into the moral narratives around vaccine hesitancy, analysing a large volume of user-generated data. 
Our findings show that people centre their argumentation around the moral value of liberty, care, and authority.
When focusing on the argumentation of each side, we notice that vaccine advocates prioritise the need to comply with authorities to protect society as a whole, while sceptics favour the value of liberty, expressed as freedom of choice on whether to adhere or not to the vaccination.
These insights can be constructive since public health communication campaigns commonly develop their narratives around care and the shared responsibility of protecting the most vulnerable in our community.
However, such messaging does not reflect the concerns of those sceptical about vaccinations who prioritise notions of freedom of choice and alternative health remedies.
We hope these findings will inform policymakers and practitioners to craft communication campaigns to bridge the mistrust between health care and the public~\cite{dhaliwal2020antivaccine,dube2015strategies}.

\section{Related Work}
Social science research has shown that vaccination decision-making should be understood in a broader socio-cultural context~\cite{dube2013vaccine}.
Vaccine hesitancy has a long history, with the role of traditional media and later of the Internet to offer an opportunity for vocal anti-vaccination activists to diffuse their message~\cite{zimmerman2005vaccine}.
SM platforms have further magnified this phenomenon, allowing individuals to rapidly create and share content globally without editorial oversight~\cite{puri2020social}. 
SM users are subject to confirmation bias and selective exposure, interacting only with like-minded peers in the so-called echo-chamber effect~\cite{cinelli2021echo}.

Such phenomena can be particularly concerning since often they proliferate misinformation while have been noticed both on the Twitter platform~\cite{cossard2020falling} and Facebook~\cite{schmidt2018polarization}\footnote{In the light of these findings, Facebook took several actions to tackle fake news decreasing their proliferation by adding a quality control \url{https://about.fb.com/news/2019/03/combatting-vaccine-misinformation/}}.
Even more concerning findings with broad implications for public health messaging were brought to light regarding the advertising communication campaigns placed on the Facebook platform by different stakeholders, including politicians, entrepreneurs, and public health authorities, in an intense competition for the audience's attention~\cite{mejova2020covid}.

Despite this, not all online communication around vaccination is harmful; a study on a parenting forum showed that users were eager to exchange opinions and learn from each other~\cite{betti2021detecting}. 
Applying natural language processing techniques to a corpus of more than one million comments from BabyCenter US, a popular parenting forum, the authors showed that hesitant parents were mainly influenced by personal and peer environments' adverse reactions. At the same time, they were more likely to be interested in alternative medicine and natural lifestyles, suggesting a broader distrust of healthcare providers and mainstream medicine.

Focusing on the Facebook Platform, Klimiuk et al.~\cite{klimiuk2021vaccine} manually categorised $\approx 20,000$ comments on vaccination in Poland as per the main categories of hesitancy. In decreasing order of importance, they ranked conspiracy theories, falsehoods, concerns regarding the safety of vaccines, and violations of the human right of liberty, presenting the compulsory vaccination as an act of ``totalitarianism''.
A limitation of this study is that the authors considered only one Facebook Page for three months.
Similar clusters were obtained by Hoffman et al.~\cite{hoffman2019s} from the manual analysis of Facebook status posts on a sample of $197$ individuals in the US. 
Kalimeri et al.~\cite{kalimeri2019human} analysed the opinions of a cohort of approximately 2,000 participants from Italy who were following Facebook Pages about vaccination via online administered questionnaires. Their findings showed that those more sceptical about vaccine uptake showed more anti-authoritarian attitudes; in particular, they did not trust the presidency of the country and the current government while also expressing anti-European opinions. Moreover, they vigorously defend the traditional moral values of Italian society and, in particular, the religion while explicitly stating that newer lifestyle leads to the decline of civilisation.

\begin{table}[h]
\renewcommand{\arraystretch}{1.2}\centering
\caption{\label{statistics}Dataset statistics. The last row points out the number of comments effectively used to train and evaluate our models, which were filtered to contain at least five words, excluding mentions.}
\begin{tabular}{ccc}
 & {PV Pages} & {AV pages} \\
\toprule

Pages & $101$ & $85$ \\
Original Posts & $52,894$ & $24,615$ \\
Original Comments & $215,341$ & $391,764$  \\
Filtered Comments & $170,954$ & $286,111$ \\
\bottomrule
\end{tabular}
\end{table}

Such findings are emerging from more than online and social media platform data.  
Studies employing traditionally administered surveys reach the same conclusions. 
In particular, Rossen et al.~\cite{rossen2019accepters} administered an online questionnaire to $296$ individuals in Australia. The authors showed that vaccine-hesitant individuals valued significantly fewer concepts of authority and more notions of purity, liberty and fairness.
Previously, Amin et al.~\cite{amin2017association} surveyed 1,000 parents in the US, showing that the higher the vaccine hesitancy, the higher the level of purity and liberty moral foundations. 
Interestingly, they also showed that the harm/care foundation, intensively addressed in traditional pro-vaccination campaigns, did not significantly impact vaccine hesitancy. 

Moral values detection from user-generated text is not a trivial problem since values are often only implicitly signalled in language.
The majority of predictive models developed to address this task are trained on easily accessible Twitter data~\cite{araque2020moralstrength,Mooijman2018}, while they are often based on computational linguistics and lexicon-based approaches~\cite{graham2009liberals,araque2020moralstrength,araque2022libertymfd}. 
Except for a preliminary study by Araque et al.~\cite{araque2021liberty,araque2022libertymfd}, none of the well-established lexicons includes the ``Liberty'' moral foundation, which plays a central role in the vaccination debate according to the related literature. \KK{note to self: give more info}
Here, contributing to the current state of the art, we propose an entirely data-driven approach based on neural networks to infer the moral narratives in user-generated text. 


\section{Data Collection}

 \begin{table}[h]\renewcommand{\arraystretch}{1.2}\centering
 \caption{\label{tab:table_annotations} Annotation of Moral values. The content of each comment was annotated with respect to the moral value expressed as Care/Harm, Fairness/Cheating, Loyalty/Betrayal, Authority/Subversion, and Purity/Degradation,Liberty/Oppression, or ``Non moral''; Virtue and Vice denote positive and negative polarity in the foundation spectrum.  Note that a comment may express more than one moral values. }
 \begin{tabular}{lccc}
 & Presence  { ~ } & {Virtue } { ~ } & {Vice } \\
\toprule
 Authority { ~ } & 314 & 110 & 204\\
 Liberty  { ~ } & 205 & 140 & 65\\
 Loyalty  { ~ } & 78 & 40 & 38\\
 Care  { ~ } & 489 & 357 & 132 \\
 Fairness  { ~ } & 297 & 174 & 123\\
 Purity  { ~ } & 192 & 80 & 112\\
 Non-moral  { ~ } & 353 & - & -\\
 Non-relevant  { ~ } & 2883 & - & -\\
 \bottomrule
 \end{tabular}
 \end{table}
 
Replicating the methodology of \cite{schmidt2018polarization}, we collected data from Facebook Pages in the period Jan 2012-Jun 2019, using the Facebook Graph API~\cite{FBapi}.
We queried the API for public Pages	containing	the	keywords	``vaccine'', ``vaccines'', or ``vaccination'' in their name. Then we excluded from the analysis Facebook pages that were not in the English language and were not relevant to the topic of vaccination (e.g., the Facebook page of the music band ``The	Vaccines''). The remaining pages were manually categorised according to their content into supportive (hereafter, PV, for brevity) or contrary (hereafter, AV) to vaccination.  
We also collected all public posts and comments from each Page.
Overall, we obtained 607,105 comments in $186$ ``Pro'' and ``Anti''\footnote{Here, we conventionally used the term ``Anti'' as an abbreviation to refer to the vaccine-hesitant population.} vaccination pages. After removing comments with less than five words, our final dataset consisted of approximately 450,000 comments. A summary of the dataset is shown in the upper part of Table~\ref{statistics}.

Since vaccination is a highly controversial topic, and despite the moderation of Facebook Pages, the published posts may bring up discussions where both sides are expressed regardless of the primary stance of the Page.
Hence, we annotated 4,498 comments independently of the affiliation of the Page on which they were written. We employed nine annotators, all familiar with the moral foundation's theory. 
For each comment, we annotated it as supporting (pro-), opposing (anti-) or not taking any stance (non-relevant) according to its position towards vaccination. For those comments that did take a stance towards vaccination, we further annotated the expression of moral values, if any, indicating the ``vice'' or the ``virtue'' of the foundation. 
Table~\ref{tab:table_annotations} reports the distribution of the annotated comments. We assessed the inter-annotator agreement via Cohen's kappa coefficient, obtaining .78 (substantial agreement) for the first part of the annotation concerning the relevance to vaccination and .32 (fair agreement) for the moral value expressed in the comment. The latter shows the difficulty of the task also for human annotators.

A limitation of this study is that we need to have information about the author of the comments; hence, we cannot draw any conclusion about the adequate size of each community or the users' engagement in the discussion.

\section{Methods}\label{sec:methods}

\begin{figure*}[ht]
\centering
    \begin{subfigure}{0.8\textwidth}
      \includegraphics[width=\linewidth]{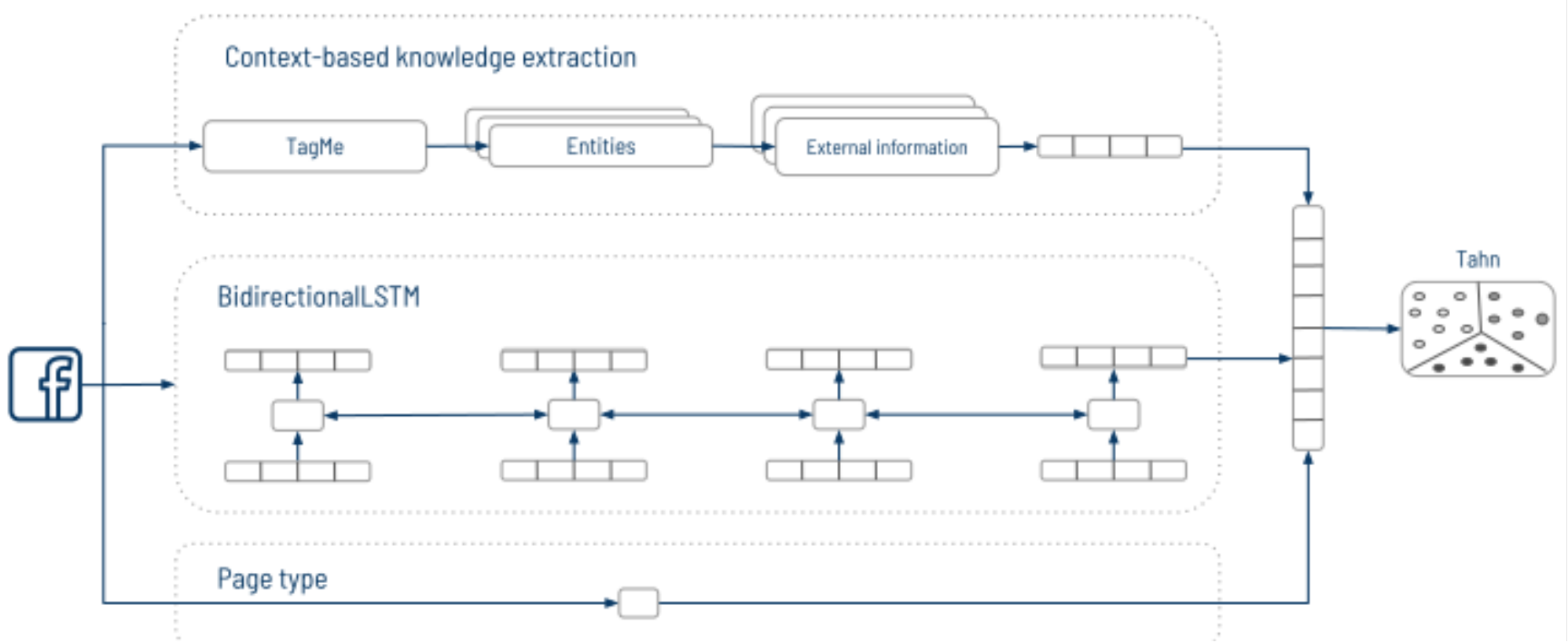}
        \caption{\label{arch_comment}Deep network architecture used for predicting comment types.}
    \end{subfigure}
    
    \begin{subfigure}{0.8\textwidth}
      \includegraphics[width=\linewidth]{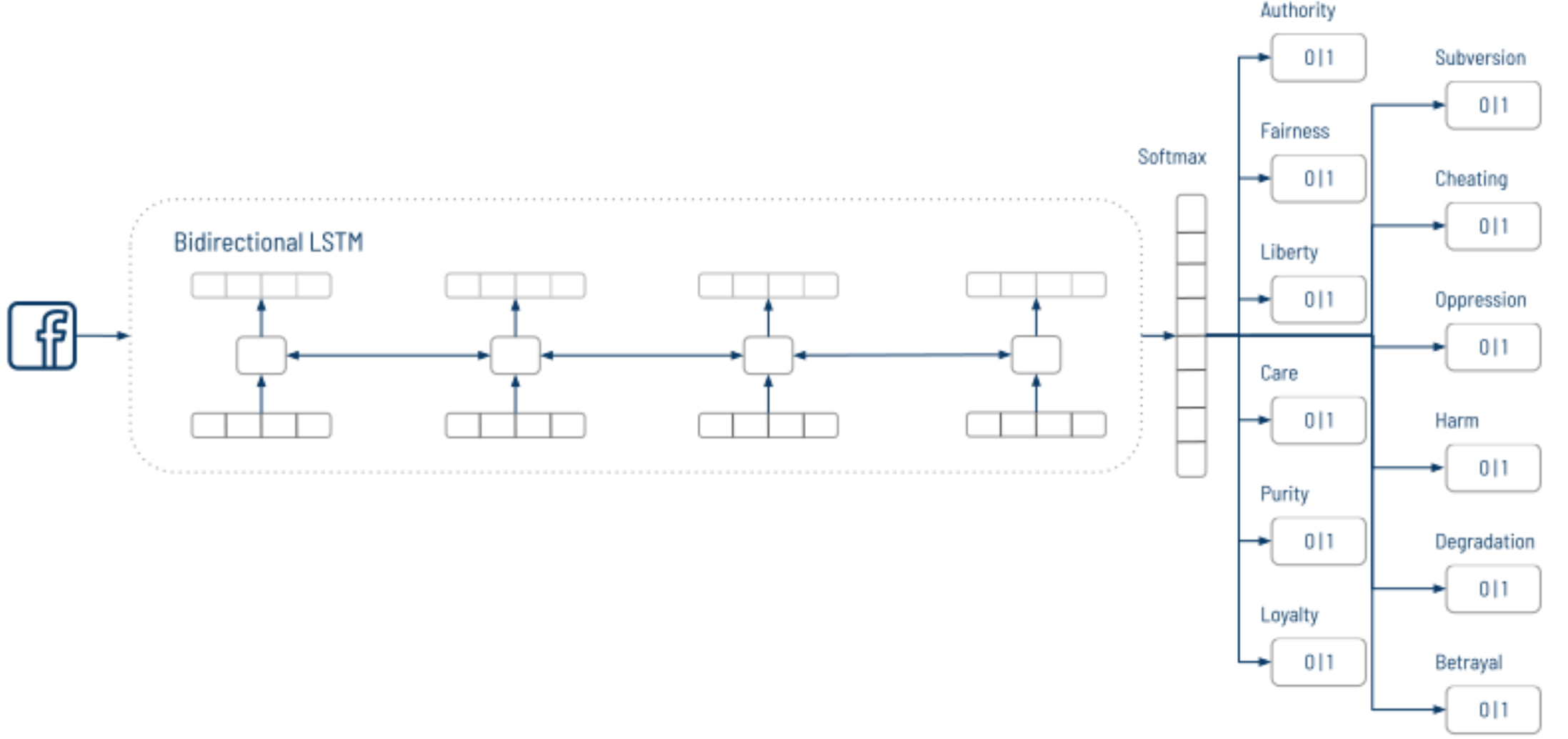}
     \caption{\label{arch_morals}Deep network architecture used for predicting moral foundations.}
    \end{subfigure}
    \caption{\label{fig:arch} Overview of the deep network architectures employed for prediction of relevance to vaccination~\ref{arch_comment} and moral values~\ref{arch_morals}. For relevance prediction~\ref{arch_comment}, the embeddings of each comment go through two parallel pipelines: one processing the text with a bidirectional LSTM, and the other applying entity recognition through TagMe.
    For moral values prediction~\ref{arch_morals}, the embeddings of each comment enter a bidirectional LSTM, and the final layer predicts each moral's virtue and vice using a softmax activation function.
    }
\end{figure*}

The complexity of the reasoning and the background knowledge often entailed in the lengthy comments posted on the Facebook Platform render the above methodologies inadequate.
Here, we propose an approach based on a Recurrent Neural Network-based classifier with
long short-term memory (LSTM)\footnote{The architecture was implemented in Keras~\cite{chollet2015keras}.} inspired by the work of Lin et al.~\cite{Lin2018}.

Initially, we aim to identify the comments that indeed argue about vaccination. To do so, we build a Relevance Classifier which predicts whether a specific comment is explicitly favouring (``Pro''), hesitant (``Anti''), or not expressing a position (Non-relevant, ``NR'') towards vaccination.
The Relevance Classifier is composed of three parallel stages (see Fig.~\ref{arch_comment}): an LSTM neural network, a named entity recognition stage based on TagMe~\cite{ferragina2010tagme} and a page class stage. 
Then, we further classify the comments relevant to vaccination concerning the appeal to morality in terms of the presence of a moral foundation (e.g. ``Care-Presence'' vs ``No Care-Presence'') and its polarity (e.g. ``Care-Virtue'', ``Care-Vice''). The structure of the Moral Classifier consists of a word embedding component and an LSTM neural network (see Fig.~\ref{arch_morals}).
We train the relevance and moral classifiers on the annotated subset of 
comments; then, these predictive models are employed to infer the unlabelled part of the dataset.
 
In detail, for the Relevance component, we apply basic natural language pre-processing techniques (i.e. stemming, removing punctuation and stopwords, tokenisation) and then employ the 100-dimensional pre-trained Glove embeddings~\cite{pennington2014glove} model to obtain the word representation.
To incorporate background knowledge, we apply entity linking to associate mentions with their referent entities, using  TagMe~\cite{ferragina2010tagme}. 
TagMe is a powerful tool that identifies on-the-fly meaningful substrings (called ``spots'') in an unstructured text and links each of them to a relevant Wikipedia page efficiently and effectively. 
TagMe provides a confidence score of $\rho$, an estimate of the annotation's ``goodness'' concerning the other entities of the input text. For this study, we threshold this confidence score to $\rho \geq 0.1$ and utilise them as additional features to improve the prediction.
The bidirectional LSTM takes as input the sequence of word embeddings 
$\{w1, w2, ..., wn\}$ in a text and output hidden
states $\{h1, h2, ..., hn\}$;  the last output $hn$ is then passed to the fully connected layer.
Finally, we concatenate them with the vectors from the background knowledge and add a final dense layer with a $\tanh$ activation function.
The training dataset was built from a subset sample of the complete annotations to have a balanced dataset among the Pro, Anti and NR classes; the training was thus performed on 2100 comments, 700 for each class.
To prevent overfitting, we apply dropout to outputs of the embedding, LSTM, and fully connected layers, while we also perform 10-fold cross-validation. 

For the comments in our dataset classified to express a ``Pro'' or an ``Anti'' vaccination opinion, we trained one LSTM model to predict each moral foundation's presence and polarity. 
The Moral component follows the same pre-processing and embedding approach~\cite{pennington2014glove} model to obtain the word representation.
Similarly, these are fed to the bidirectional LSTM. In the case of polarity prediction, we perform 5-fold cross-validation due to the limited amount of labelled data per moral class (see Table~\ref{tab:table_annotations}).
We report the Area Under the Receiver Operating Characteristic curve (AUROC) for all steps for each classifier.
Figure~\ref{arch_morals} depicts the architecture of this design.

\section{Results and Discussion}

\begin{figure*}[ht]
    \centering
    \includegraphics[width=0.955\textwidth]{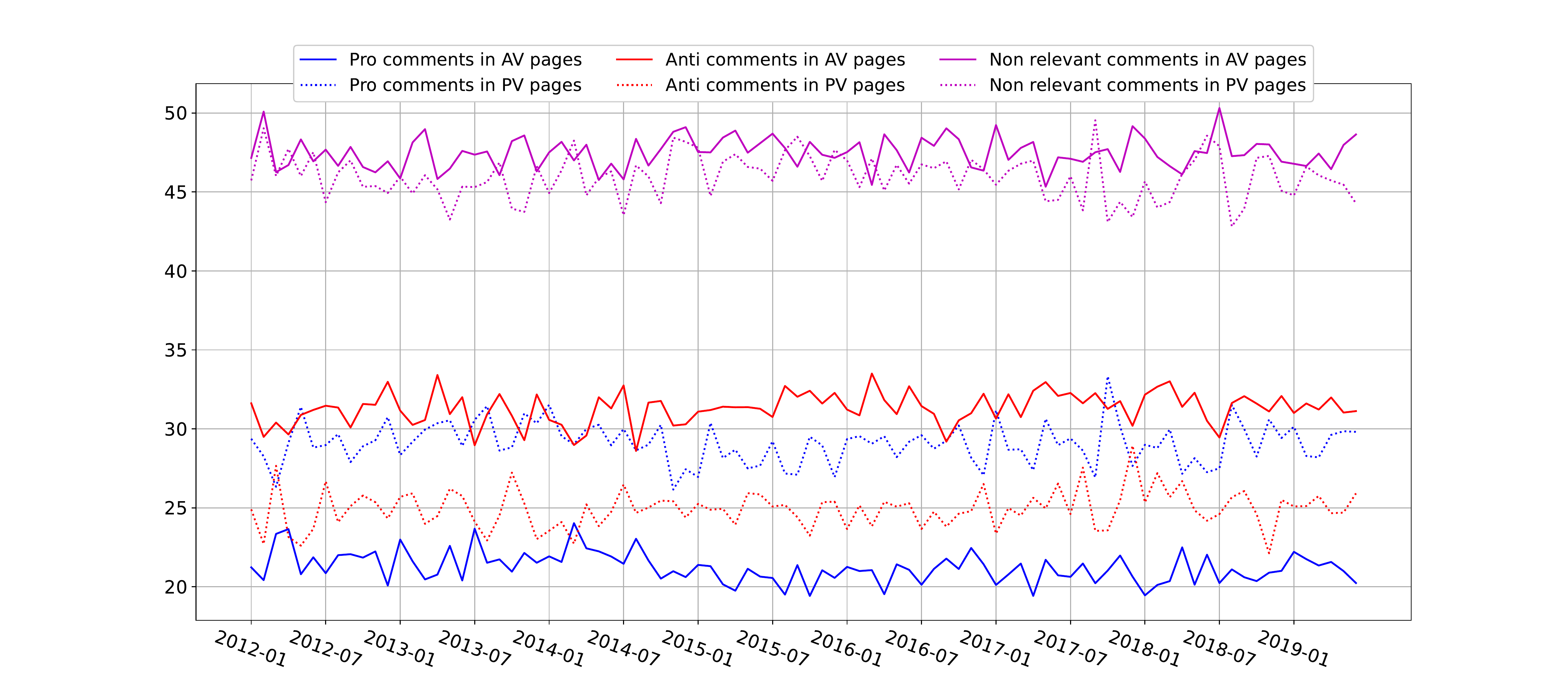}
    \caption{Monthly percentages of comments classified as (PRO, ANTI, NR)-vax in Pro-vax and Anti-vax pages during time.}
    \label{fig:pna_stats_timeseries}
\end{figure*}

\KK{replace figure\ref{pna_models} with table~\ref{rocs_pv_av}}

\begin{table}[h]\centering
\renewcommand{\arraystretch}{1.2}
\caption{\label{rocs_pv_av} Cross-validated performance (AUROC) of different models for the prediction of a comment's position: (PRO, ANTI, NR)-vax. Regression Model, LSTM branch of the model and the full model as described in Section~\ref{sec:methods}. The Regression Model serves as a baseline model.}
\begin{tabular}{lrrr}
{} &  Regression Model &  LSTM branch &  LSTM full \\
\toprule
\emph{PV comments}  &  $.51 \pm .03$   &  $.63  \pm .04$ & $ .71 \pm .06$\\
\emph{AV comments} & $.51  \pm .0$4  &  $.80 \pm .03$ & $.82 \pm .02$\\
\emph{NR comments}  &  $.50  \pm .04$ &  $.71 \pm .04$  & $.80 \pm .05$\\
\bottomrule
\end{tabular}
\end{table}
\begin{table}[h]\centering
\renewcommand{\arraystretch}{1.2}
\caption{\label{table_pna_percentages}Percentages of comments classified as Pro-Vax, Anti-Vax, or Non-Relevant (NR) in regard to vaccination stances. Second and third column express the percentage of comments expressing these stances in Pro-Vax and Anti-Vax Facebook pages in our dataset.}
\begin{tabular}{lrrr}
{} &  All pages &  PV pages &  AV pages  \\
\toprule
\emph{PV comments}  &      24.18\% &      29.13\% &       21.22\% \\
\emph{AV comments} &      28.94\% &      24.91\% &       31.35\% \\
\emph{NR comments}  &      46.88\% &      45.96\% &       47.42\%\\
\bottomrule
\end{tabular}
\end{table}

Initially, from our data collection, we noticed that overall the AV pages are fewer, with almost half of the posts, which triggered almost double the comments concerning the PV pages. This observation is confirmed by the findings reported in the currently related literature regarding the vaccine sceptics, who appear to be an extremely vocal minority~\cite{cossard2020falling}.

\textbf{Stance Classification.} Moving to the relevance classification, our model accurately predicted the stance towards vaccination with an average accuracy of  $.76$ on a 10-fold cross-validation scheme. We conventionally refer to the AUROC curve (Area Under the Receiver Operating Characteristic) metric as ``accuracy''. We opted for this metric to evaluate our framework due to its stability in heavily imbalanced classification tasks like ours. 
This is an encouraging finding showing that the model can distinguish whether an unseen comment is supportive (``Pro'') or sceptical (``Anti'') towards vaccination or whether it does not take any position. 
When applying this model to predict the comments' stance on the entire dataset, we notice that approximately half the comments do not express a firm position on the debate. The vaccine sceptic comments are more frequent than the supportive ones (see Table ~\ref{table_pna_percentages}), confirming the characterisation of the vaccine sceptics as a vocal minority. However,  we find that despite the persistent content moderation observed on the Facebook platform leading to echo-chamber effects~\cite{schmidt2018polarization}, in both PV and AV pages, there exists an appropriate amount of comments expressing the opposite stance, which suggests that there is significant interaction and exposure to different opinions within both communities~\cite{betti2021detecting}. Figure ~\ref{fig:pna_stats_timeseries} shows that the evolution in time of the percentages for pro- anti- and non-relevant comments remain stable. 

As a baseline for comparison, we considered two approaches: a benchmark model implemented as a Logistic Regression classifier trained on a bag-of-words representation for each comment. Secondly, an ablation study where the LSTM model is trained only on text features. This latter evaluation step accounts for the importance of the annotations and page information for the final prediction task. \KK{note to self: improve this}
Table~\ref{rocs_pv_av} shows that our architecture (third column) significantly outperforms the Logistic Regression model (first column) as well as the simple LSTM architecture (second column), highlighting the importance of the TagMe and pages features for understanding the stance towards vaccination.

\textbf{Moral Presence Classification.}
In the second step of the classification, we detect the presence of each moral foundation in the vaccine-related comments. Overall, from the entire dataset, 107,712 comments have been classified as vaccine supportive, while 128,939 as hesitant; the remaining ones, around 210,000, have instead been classified as non-relevant to the topic of vaccination.
Table~\ref{table_rocs} depicts the average AUROC for each foundation trained and tested on the labelled data with a 10-fold cross-validation scheme. 
We notice that the loyalty foundation is the hardest to predict; this is a straightforward limitation of the annotated sample of comments since there the Loyalty was the rarest foundation to be detected, and hence the training samples were few. \KK{check if I can say something about the annotators' agreement per foundation.}
Interestingly, Lin et al.~\cite{Lin2018} also report LoyaltyLoyalty as the most challenging foundation to predict, despite the higher availability of labelled training data.
The Liberty and care moral foundations were the ones to be predicted most accurately with .71 and .68, respectively (see Table~\ref{table_rocs}).
Especially for the case of the Liberty foundation, this finding shows that despite the limited annotated instances for the other foundations (205 for Liberty versus 489 for Care), the content is quite consistent, achieving the highest accuracy.

\textbf{Moral Polarity Classification.} 
Finally, we trained a multi-target classifier with twelve binary targets, one for each foundation and polarity. 
Table~\ref{table_rocs} (Virtue and Vice columns) depict the AUROC scores
for the Virtue and the Vice polarities of each foundation, respectively.
Here we find that Authority is the best-performing one, with an AUROC of $.65$ on either polarity. Overall, the performance of this model is poorer than that of the model predicting the presence of each moral foundation, probably due to the lack of training data (see Table~\ref{tab:table_annotations})\KK{ i don't understand the rest of the sentence} and the fact that the presence prediction model puts together the annotations regarding Virtue and vice of each foundation, thus gathering more data.

 \begin{table}[ht]\centering
 \renewcommand{\arraystretch}{1.2}
 \caption{\label{table_rocs}Average and Standard Deviation of the LSTM predictors accuracy for each of the six moral foundations, expressed in terms of the area under the receiver operating characteristic (AUROC).}

 \begin{tabular}{l ccc}
 &
 {Presence    {~ }} & {Moral Virtue} { ~ } & {Moral Vice } \\
 & &  (Positive Polarity) &  (Negative Polarity) \\
 \toprule
 Authority { ~ } & $.61 \pm .07$ & .65 & .65\\
 Liberty  { ~ } & $.71 \pm .08$ & .58 & .61\\
 Loyalty  { ~ } & $.63 \pm .12$ & .71 & .61\\
 Care  { ~ } & $.68 \pm .05$ & .62 & .55 \\
 Fairness  { ~ } & $.60 \pm .06$ & .56 & .60\\
 Purity  { ~ } & $.61 \pm .08$ & .49 & .57\\
 \bottomrule
\end{tabular}
\end{table}

\begin{table}[h]\centering
\renewcommand{\arraystretch}{1.2}
\caption{\label{table_presence_valence}Measures of moral content predicted by the moral values classifier. The first three columns represent the percentage of comments that exhibit the presence of a given moral (either positive or negative) in comments of different type: Pro-vax (PV), Anti-vax (AV), and Non-relevant (NR). Ratios in the three remaining columns represent, among those comments, the fraction of positive ones (virtue) over negative ones (vice) in relation to that moral value.}

\begin{tabular}{l ccc|ccc}
{} & \multicolumn{3}{c}{Occurrences} & \multicolumn{3}{c}{Virtue vs Vice Ratios} \\

\midrule
Authority &             9.40\%&             30.76\% &              36.23\% &       0.34 &        0.63 &         0.17  \\
Liberty   &             8.58\%  &             12.50\%  &              16.07\% &       0.72 &        1.49 &         2.20 \\
Loyalty  &             9.38\% &              9.07\%  &               7.75\% &       0.08 &        0.29 &         0.48\\
Care   &             8.98\% &             26.44\%  &              36.38\%    &       1.14 &        2.82 &         2.02  \\
Fairness  &            25.58\%  &             24.55\%  &              21.07\%   &       0.05 &        0.46 &         0.89 \\

Purity  &            12.67\%  &             21.33\%  &              22.87\%    &       0.34 &        0.50 &         1.03 \\
\bottomrule
{} &  \emph{NR} &   \emph{PV} &   \emph{AV} &  \emph{NR} &   \emph{PV} &   \emph{AV} \\
\end{tabular}
\end{table}

To assess the moral narratives in how people reason about vaccination, we confronted the moral presence of messages that support vaccination as opposed to the hesitant ones (see Table~\ref{table_presence_valence}). Note that a comment may express more than one moral foundation or be non-moral: from the results of the classifier over all the 12 combinations across moral foundations and polarities, we indeed find that around $\sim 32\%$ of the comments are assigned more than one moral value, $\sim 35\%$ just one, and $\sim 33 \%$ are assigned no moral foundation at all.

As a measure of the leaning of each group towards the extremes of the moral axis, we evaluated the ratio of comments classified as positive and negative for each foundation. This is:
\begin{equation}
    VVR(M, S)=\frac{P(Virtue|M,S)}{P(Vice|M,S)}
\end{equation}
where $M$ represents a moral foundation and $S$ is the stance towards vaccination. A high value of this ratio tells us that group $S$ highly values foundation $M$ as a virtue. Thus, comparing these ratios between $Pro$ and $Anti$ comments is a measure of how vaccination support and hesitancy correlate to moral values.

\begin{figure*}[ht]
    \centering
    \includegraphics[width=\textwidth]{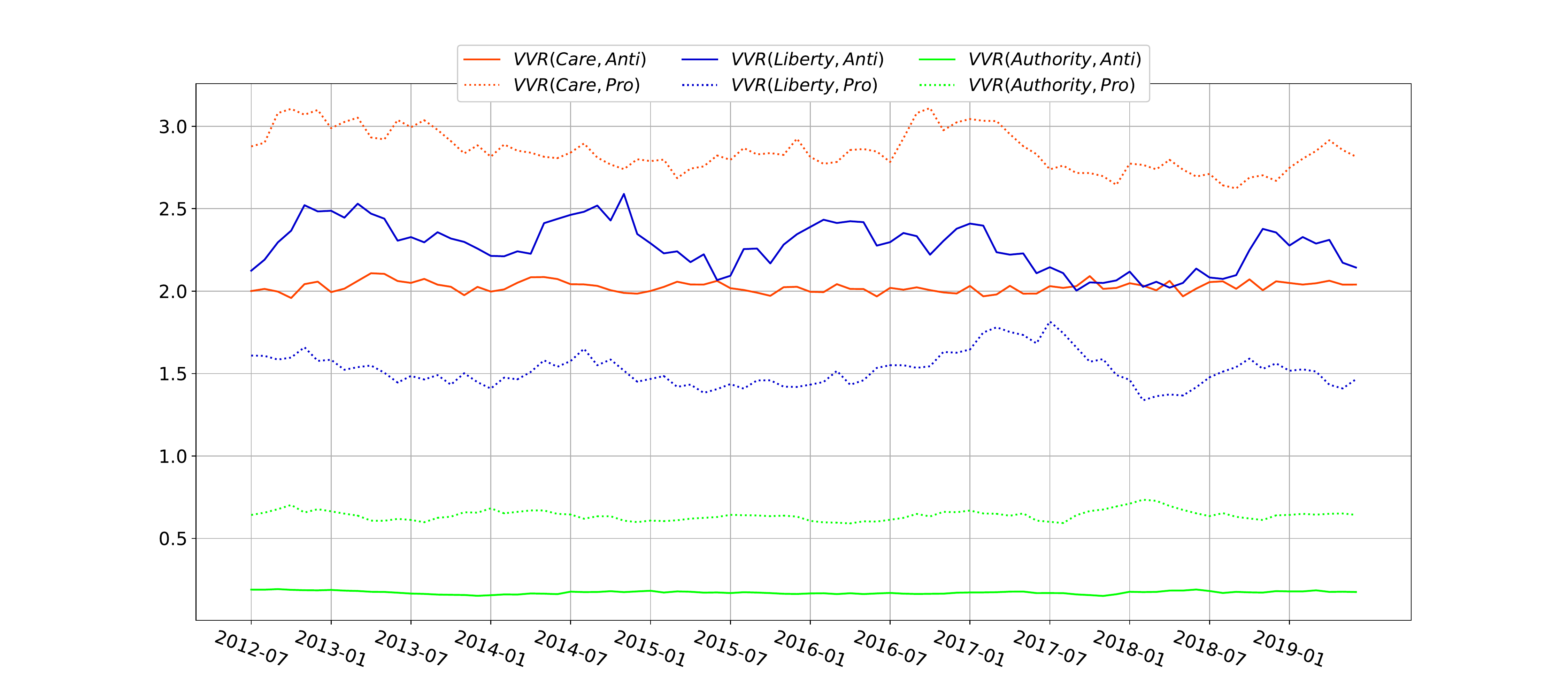}
    \caption{VVR (Virtue vs Vice Ratio) over time for the Care, Liberty and Authority foundations expressed in predicted Pro-vax and Anti-vax comments. To denoise we applied a moving average filter per month with a six month window frame.}
    \label{fig:VVR_comparison}
\end{figure*}

The analysis of these ratios for the different morals and groups is shown in Table~\ref{table_presence_valence}. We observe that comments classified as opposing vaccination express more intensely the moral value of Liberty (Virtue) ($VVR$ of $2.20$) in comparison with those supporting vaccination ($VVR$ of $1.49$). 
Moreover, vaccine-hesitant individuals are firmly positioned against the authoritarian actions of those in charge ($VVR$ of $0.17$).
Such findings confirm the current state-of-the-art literature~\cite{amin2017association,kalimeri2019human,betti2021detecting}, which underlines the importance that the vaccine-hesitant community gives to notions related to freedom of choice.
Focusing on the liberty, care and authority foundations, Figure~\ref{fig:VVR_comparison} depicts the Virtue vs Vice Ratio (VVR) over time. We notice a stable and consistent difference in moral framing, with the vaccine advocates expressing more concerns about the protection of others while the hesitant tend to focus more on the freedom of choice and self-determination argumentation. We also find that comments supporting vaccination express more familiar notions of the Vice spectrum of the Purity foundation; such narratives often criticise natural remedies and holistic alternatives to vaccination in line with the  literature~\cite{kata2010postmodern}. 

The importance of these findings lies in the insights provided into the moral narratives and argumentation of vaccine advocates and hesitant. In critical social issues such as vaccine hesitancy, mutual understanding of the other's side argumentation is fundamental to crafting communication campaigns addressing these specific concerns and drivers of hesitancy.

\section{Conclusions}

The drivers of vaccine hesitancy are complex; scientists and medical doctors have long been trying to map the causes and dependencies of that factors~\cite {DUBE2016,kata2010postmodern}. 
Even though various psychological attributes, such as personality traits, have been related to vaccine hesitancy, the role of moral values has been largely understudied.

Here, we employ the Moral Foundations Theory to assess the moral narratives in random, user-generated text from social media. 
We analysed more than 400,000 comments on Facebook pages regarding vaccination, both from the supporting and opposing sides.
In particular, we systematically assess the role of the Liberty moral foundation previously omitted.
While in the future, we aim to assess the moral narratives via bidirectional encoder representations from transformers; here, we build on the work of Lin et al.~\cite{Lin2018}. We trained a series of models based on Recurrent Neural Networks with short-term memory and entity linking. 

Here, we confirmed that the vaccine sceptic community is a vocal minority; however, despite the platform's moderation, the users are exposed to the content of like-minded peers and diverse opinions.
The deep architecture proposed is shown to outperform the baseline models while performing similarly \KK{(please check and correct here if necessary)} to existing models in moral values prediction from the user-generated text.
The main contribution of this study is the large-scale, automatic, and in-depth analysis of the moral framing of vaccine advocates and sceptics, advancing the understanding of vaccine hesitancy drivers.

Our findings confirm the importance of the liberty foundation to the discussion around vaccination and feelings of oppression from the authorities. 
Such indications are also supported by survey-based social science and public health studies; however, this is the first time such a narrative has emerged in a data-driven way from automatically analysing user-generated text.
Understanding the viewpoints of the hesitant community will hopefully lead to communication campaigns and strategies that avoid eliciting backfiring effects such as those of an obligatory vaccination act.

Interestingly, in light of the recent COVID-19 pandemic, the vaccine hesitancy drivers are likely to be rooted in the same moral narratives. 
Antivaccination campaigns built around concepts of holistic cures and natural remedies to substitute vaccines gained publicity, putting at risk the global public health~\cite{larson2018biggest}.
Such insights are critical during public health emergencies and can help prevent misinformation attempts that develop their narratives around people's concerns~\cite{mejova2020covid}.

\begin{acks}
KK and JI acknowledge financial support from the Lagrange Project of the Institute for Scientific Interchange Foundation (ISI Foundation) funded by Fondazione Cassa di Risparmio di Torino (Fondazione CRT).
\end{acks}



\end{document}